\documentclass{jps-cp}
\usepackage{txfonts} 

\newcommand{\tcite}[1]{~\cite{#1}}
\newcommand{\tref}[1]{~\ref{#1}}
\newcommand{\eref}[1]{~\eqref{#1}}

\title{Towards leading-twist $T$-odd TMD gluon distributions}

\author{Alessandro \textsc{Bacchetta}$^{1,2}$,
        Francesco Giovanni \textsc{Celiberto}$^{3,4,5}$, and
        Marco \textsc{Radici}$^{2}$}

\inst{
$^{1}$Dipartimento di Fisica, Universit\`a di Pavia, via Bassi 6, I-27100 Pavia, Italy \\
$^{2}$INFN Sezione di Pavia, via Bassi 6, I-27100 Pavia, Italy \\
$^{3}$European Centre for Theoretical Studies in Nuclear Physics and Related Areas (ECT*), I-38123 Villazzano, Trento, Italy \\
$^{4}$Fondazione Bruno Kessler (FBK), I-38123 Povo, Trento, Italy \\
$^{5}$INFN-TIFPA Trento Institute of Fundamental Physics and Applications, I-38123 Povo, Trento, Italy
}

\email{f.celiberto@ectstar.eu}

\recdate{January 25, 2021}

\abst{We present exploratory studies of the 3D proton tomography through polarized $T$-odd gluon TMDs at leading twist, obtained in a spectator-model framework. We embody in our approach a flexible parameterization for the spectator-mass spectral function, suited to catch both small- and moderate-$x$ effects. All these studies are relevant to unveil the gluon dynamics inside hadrons, which represents a core research line of studies at new-generation colliders, such as the Electron-Ion Collider, NICA-SPD, the High-Luminosity LHC, and the Forward Physics Facility.}

\kword{gluon TMDs, 3D proton tomography, Hadron Structure, factorization, QCD}

\begin{document}
\maketitle

\vspace{-1.00cm}

\section{Introduction}
\label{intro}

One of the ultimate goals of frontier researches in particle physics is unraveling the inner structure of nucleons in terms of the distribution of their constituents. 
The \emph{collinear} factorization is a well-established formalism that has collected many successes since the advent of the \emph{parton model}. 
A key role in the description of high-energy hadronic and lepto-hadronic collisions is played by the one-dimensional parton distribution functions (PDFs).
However, there are fundamental questions about the deep nature of strong interactions that are still open and whose answers go beyond the reach of a pure collinear description.
As an example, unveiling the origin of proton mass and spin requires a viewpoint stretched to a three-dimensional, \emph{tomographic} description, which is naturally provided by the so-called \emph{transverse-momentum-dependent} (TMD) factorization.

While quite solid results have been obtained both on the formal and the phenomenological side for quark TMD densities, the gluon-TMD field is still an almost uncharted territory.
In Ref.~\cite{Mulders:2000sh} a first classification of unpolarized and polarized gluon TMD distributions was afforded. It was then extended in Refs.~\cite{Meissner:2007rx,Lorce:2013pza,Boer:2016xqr}, whereas first phenomenological predictions were subsequently provided~\cite{Lu:2016vqu,Lansberg:2017dzg,Gutierrez-Reyes:2019rug,Scarpa:2019fol,COMPASS:2017ezz,DAlesio:2017rzj,DAlesio:2018rnv,DAlesio:2019qpk}.

A striking difference between TMD and collinear densities is represented by the gauge-link sensitivity. In particular, the fact that TMDs are sensitive to the transverse components of the gauge link makes them process dependent~(see Refs.~\cite{Brodsky:2002cx,Collins:2002kn,Ji:2002aa}).
Quark TMDs depend on processes through the $[+]$ and $[-]$ staple links, which determine the direction of future- and past-pointing Wilson lines, respectively. 
The gluon TMDs have a more complicated gauge-link dependence, since they are sensitive on combinations of staple links. This fact leads to a more diversified kind of \emph{modified universality}.
Two major gluon gauge links emerge: the $f\text{-type}$ and the $d\text{-type}$ ones. They are also known in the context of small-$x$ studies as Weisz\"acker--Williams and dipole structures, respectively.
The antisymmetric $f_{abc}$ QCD color structure is part of the $f$-type $T$-odd gluon-TMD correlator, whereas the symmetric $d_{abc}$ structure appears in the $d$-type $T$-odd one. This brings to a dependence of $f$-type gluon TMDs on the $[\pm,\pm]$ gauge-link combinations, while $d$-type gluon TMDs are characterized by the $[\pm,\mp]$ ones.
More intricate, box-loop gauge links appear in processes where multiple color exchanges connect both initial and final state states~\cite{Bomhof:2006dp}, thus leading however to a violation of the TMD factorization~\cite{Rogers:2013zha}.

Recent studies were done on the BFKL unintegrated gluon distribution, which is the small-$x$ and large-$p_T$ counterpart of the unpolarized gluon TMD~\cite{Bolognino:2018rhb,Bolognino:2018mlw,Bolognino:2019bko,Bolognino:2019pba,Celiberto:2019slj,Celiberto:2018muu} and, more in general, on the dynamics of strong interactions at high energies~\cite{Celiberto:2020wpk,Celiberto:2016vhn,Bolognino:2018oth,Bolognino:2019yqj,Bolognino:2019yls,Bolognino:2021niq,Celiberto:2020tmb,Celiberto:2020rxb,Bolognino:2021mrc,Celiberto:2021dzy,Celiberto:2021fdp,Celiberto:2021txb,Celiberto:2021xpm,Bolognino:2021zco,Celiberto:2021tky,Celiberto:2021fjf,Bolognino:2021hxx}.
A connection between the \emph{high-energy} and the TMD factorization was highlighted in Refs.~\cite{Nefedov:2021vvy,Hentschinski:2021lsh}.

A spectator-model calculation of quark TMDs in the proton was done in Refs.~\cite{Bacchetta:2008af,Bacchetta:2010si}. 
A comprehensive framework was recently built~\cite{Bacchetta:2020vty} (see also Refs.~\cite{Bacchetta:2021oht,Celiberto:2021zww,Bacchetta:2021lvw,Bacchetta:2021twk}) for all the $T$-even gluon TMDs at twist-2 by defining an enhanced spectator model for the parent proton to effectively catch effects coming from the high energy resummation.

In this work we report a preliminary study on the $T$-odd gluon TMDs, the $f$-type Sivers and linearity functions, which are connected to relevant single-spin asymmetries arising from the distribution of unpolarized and linearly-polarized gluons inside a transversely polarized proton.

\section{$T$-odd gluon TMDs in a spectator model}
\label{gluon_TMDs}

The spectator-model framework is based on a simple and intuitive assumption, namely that the incoming proton with mass $\cal M$ and four-momentum $\cal P$ emits a gluon having longitudinal fraction $x$, four-momentum $p$, and transverse momentum $\boldsymbol{p}_T$, and the remainders are effectively treated as an on-shell spectator particle with mass ${\cal M}_X$ and spin-1/2.
The nucleon-gluon-spectator vertex is modeled as follows
\begin{equation}
 \label{eq:form_factor}
 {\cal G}^{\, \mu} = \left( \tau_1(p^2) \, \gamma^{\, \mu} + \tau_2(p^2) \, \frac{i}{2{\cal M}} \sigma^{\, \mu\nu}p_\nu \right) \; ,
\end{equation}
the $\tau_1$ and $\tau_2$ functions being dipolar form factors in $\boldsymbol{p}_T^2$. A dipolar choice for the couplings is useful to remove gluon-propagator divergences, suppress large-$\boldsymbol{p}_T$ effects which are beyond the reach of a pure TMD description, and dampen logarithmic singularities coming from $\boldsymbol{p}_T$-integrated distributions.
All the unpolarized and polarized spectator-model $T$-even gluon TMDs at twist-2 in the proton were obtained in\tcite{Bacchetta:2020vty}. 
In that work the naive spectator-model approach was improved by allowing the spectator mass ${\cal M}_X$ to spread over a continuous range of values via a flexible spectral function suited to capture both small- and moderate-$x$ effects (see Eqs.~(16) and~(17) of Ref.~\cite{Bacchetta:2020vty}).
The model parameters encoded in the definition of the spectral function and in the spectator-model correlator were determined through a simultaneous fit of the unpolarized and helicity gluon TMD densities, $f_1^g$ and $g_1^g$, to the corresponding collinear PDF distributions obtained from {\tt NNPDF}\tcite{Ball:2017otu,Nocera:2014gqa} at the initial scale $Q_0 = 1.64$ GeV. The size of the statistical uncertainty was assessed by means of the bootstrap method.

Since the tree-level approximation for the gluon correlator does not account for the gauge link, our $T$-even TMD distributions turn out to be process-independent.
In order to generate $T$-odd structures in the gluon correlator, we need to go beyond the tree level  and include its interference with a distinct channel. Similarly to the quark TMD case, we have considered the one-gluon exchange in eikonal approximation. This diagram corresponds to the truncation at first order of the whole gauge-link operator. The main effect of this procedure is that the obtained $T$-odd functions become sensitive to gauge links, and thus process dependent. For the given $f$-type gauge link, two Sivers TMDs ($f_{1T}^\perp$) and two linearity TMDs ($h_1$) are obtained by suitably projecting the transverse part of the corresponding gluon correlator. For each pair, the two partners are connected by the following modified-universality relation
\begin{align}
 \label{eq:T_odd_TMDs_f}
 \nonumber
 f_{1T}^{\perp g \, [+,+]}(x, \boldsymbol{p}_T^2) & \; \equiv \; - \, f_{1T}^{\perp g \, [-,-]}(x, \boldsymbol{p}_T^2) \; ;
 \\
 h_1^{g \, [+,+]}(x, \boldsymbol{p}_T^2) & \; \equiv \; - \, h_1^{g \, [-,-]}(x, \boldsymbol{p}_T^2) \; .
\end{align}
In our preliminary analysis we have employed a simplified expression for the nucleon-gluon-spectator vertex, with the $\tau_2$ form factor in Eq.\eref{eq:form_factor} set to zero.
For the sake of consistency, we have fitted the model parameters to {\tt NNPDF} parametrizations by using the simplified expression for the vertex.

\begin{figure}[b]
 
 \centering

 \hspace{0.25cm}
 \includegraphics[width=0.48\textwidth]{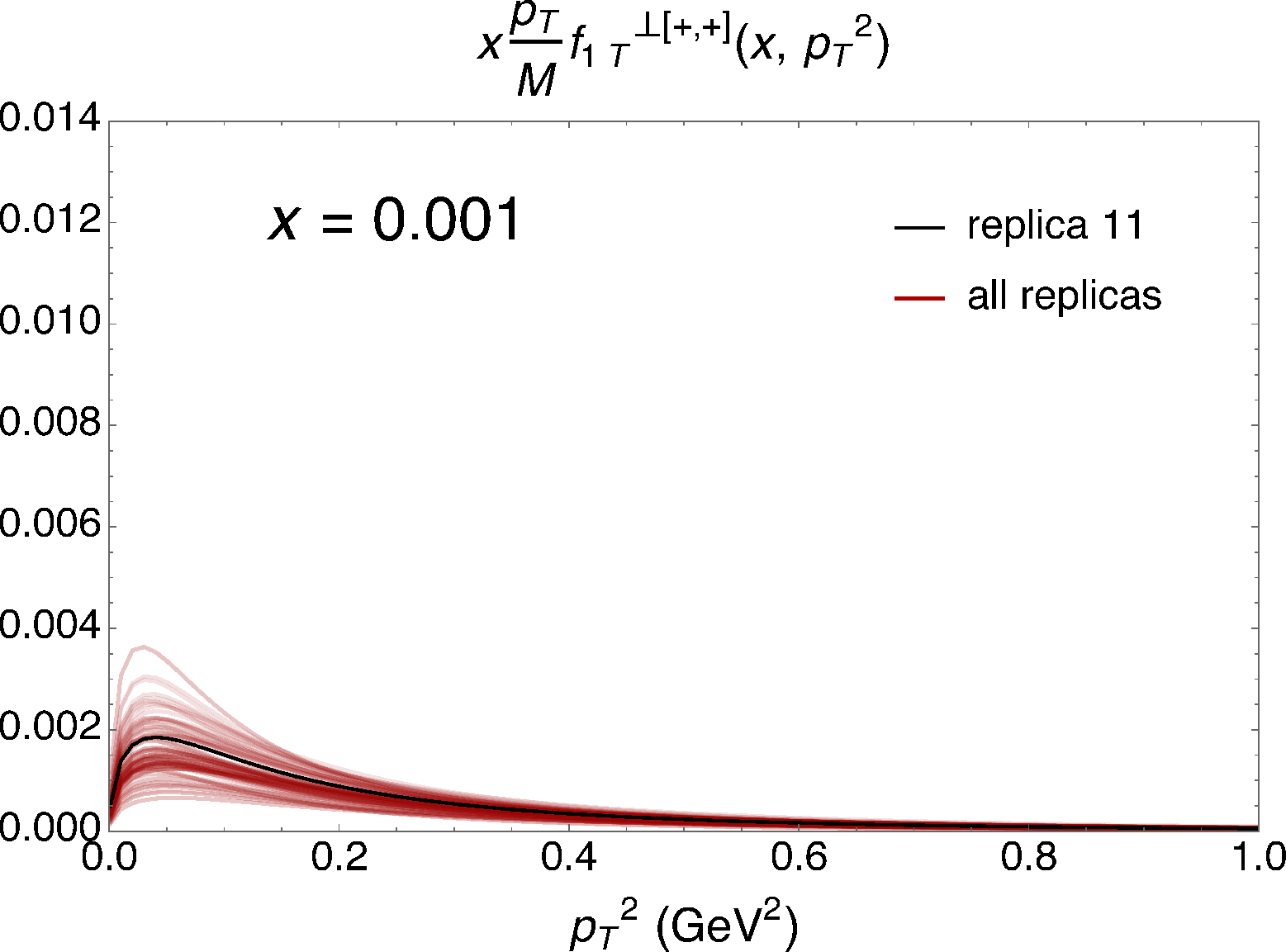}
 \includegraphics[width=0.48\textwidth]{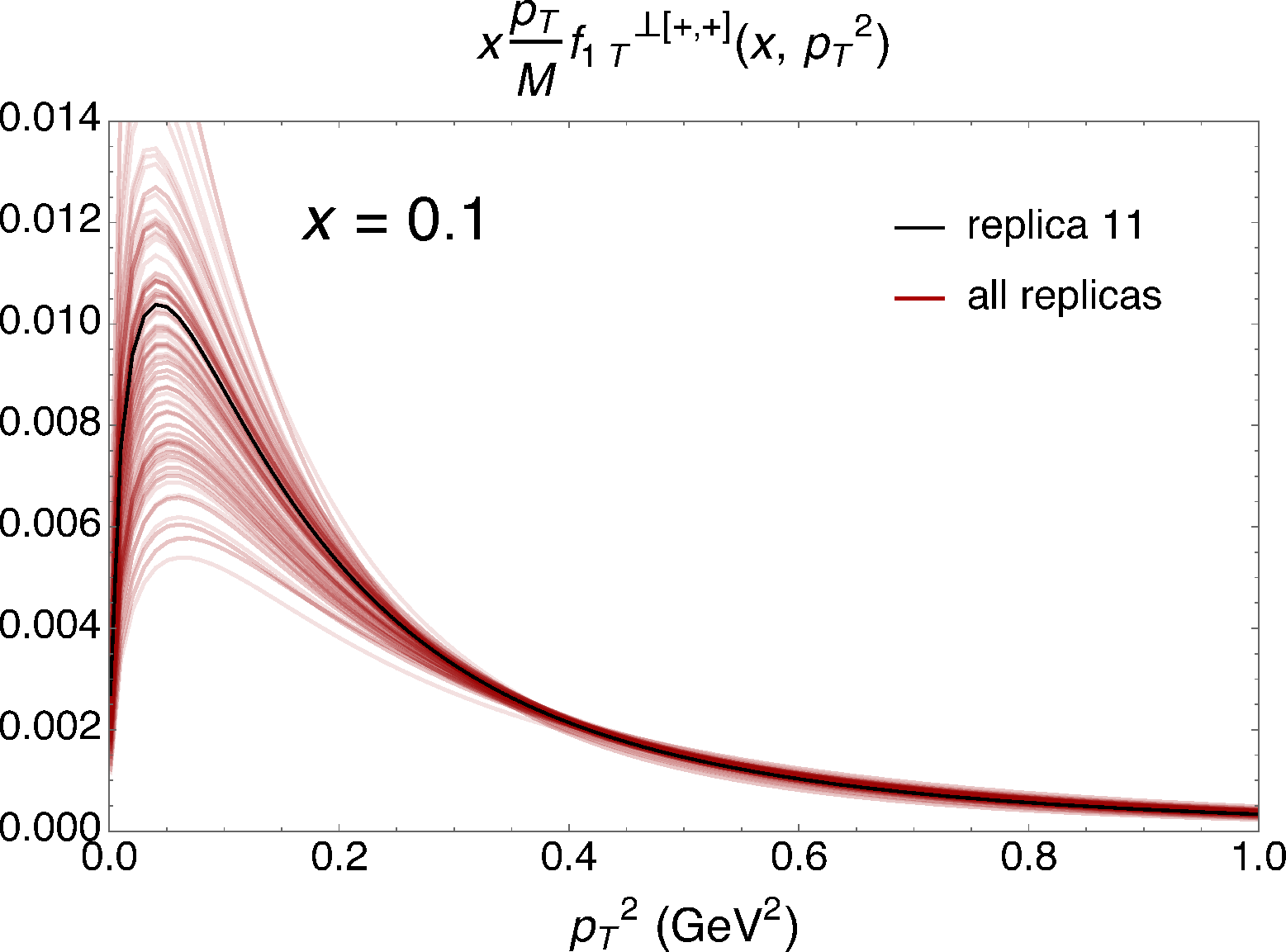}

 \vspace{0.50cm}

 \includegraphics[width=0.48\textwidth]{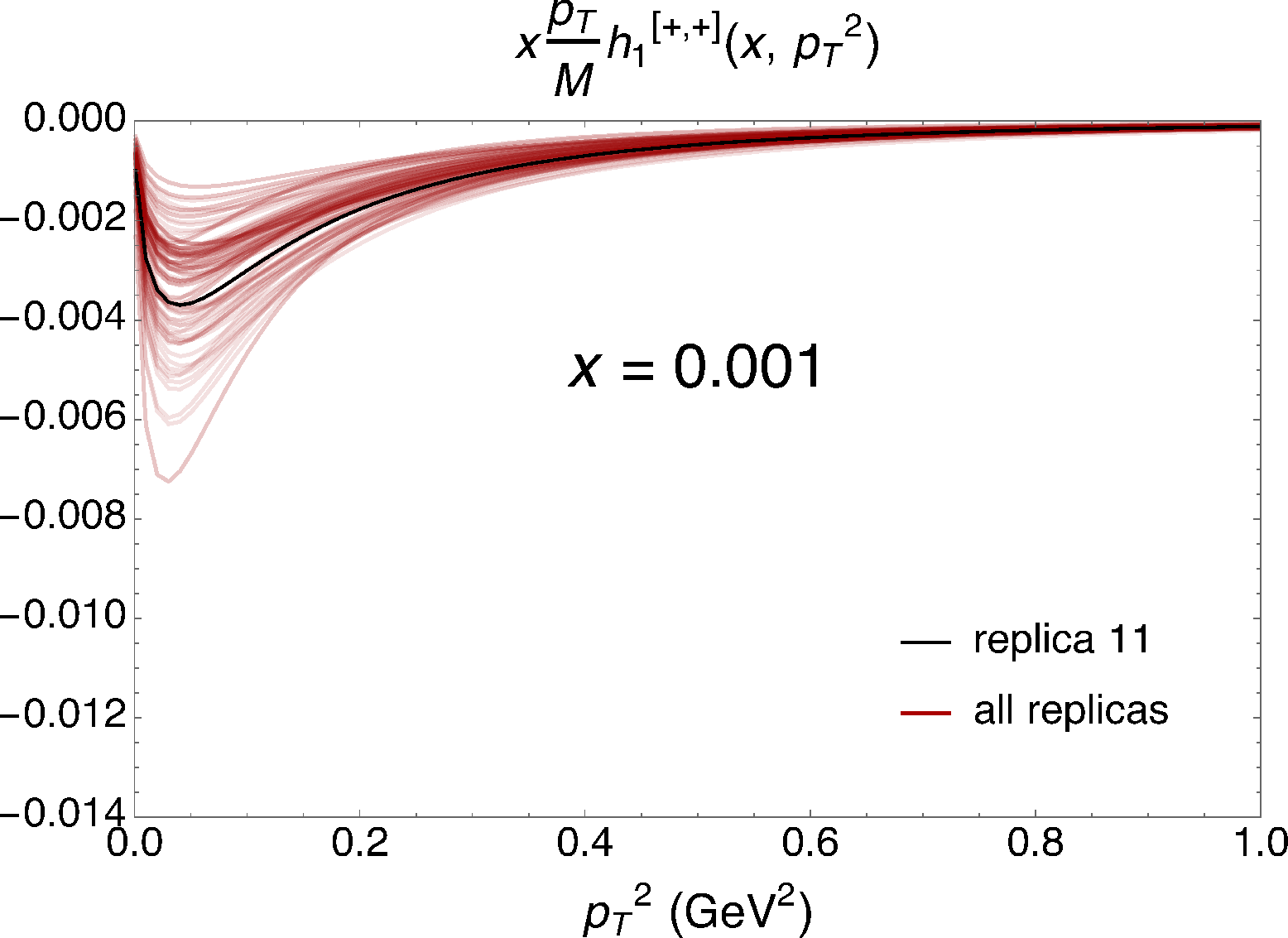}
 \hspace{0.25cm}
 \includegraphics[width=0.48\textwidth]{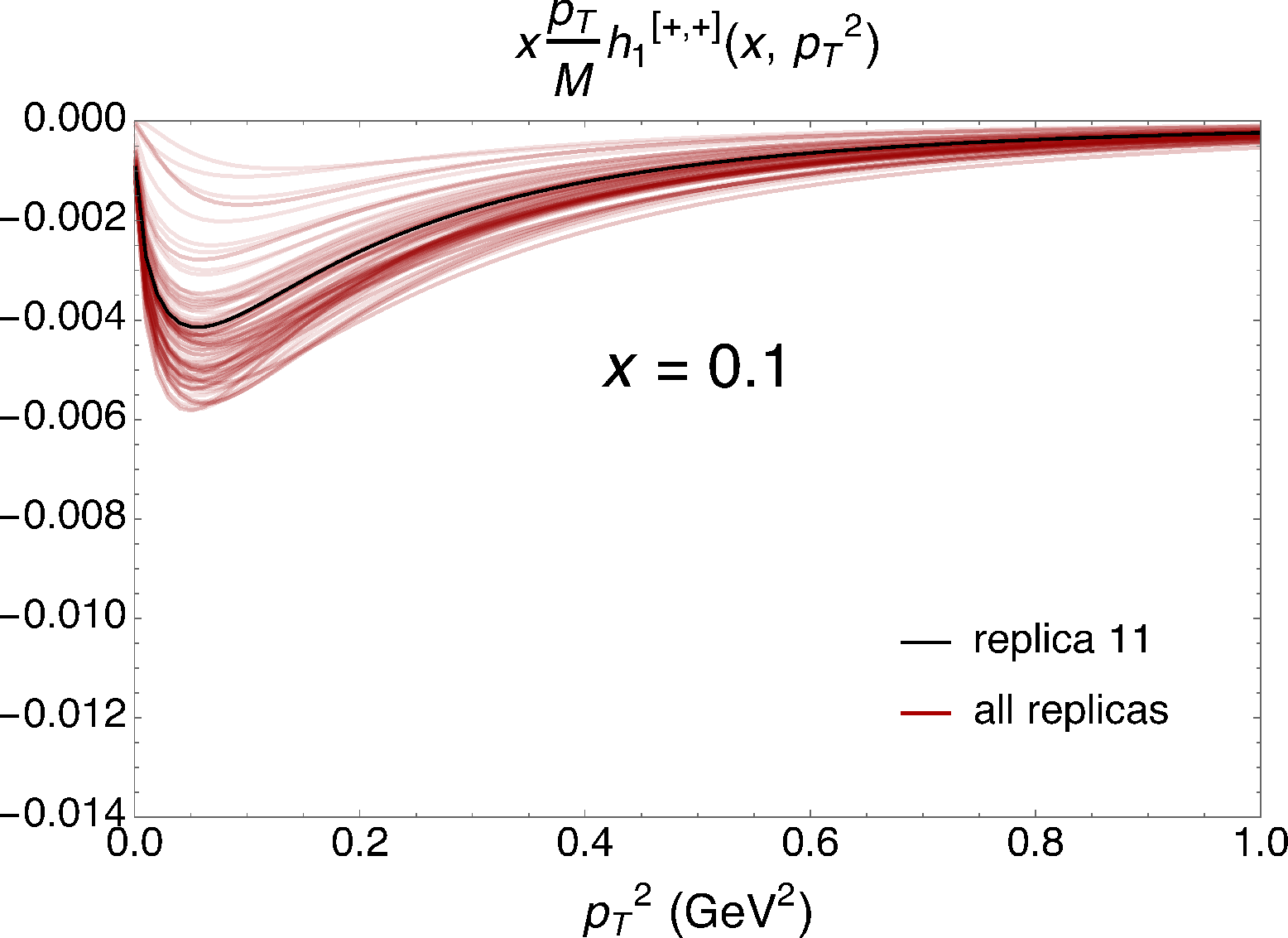}

 \caption{Transverse-momentum dependence of the $[+,+]$ Sivers (upper) and linearity (lower) densities for $x=10^{-3}$ (left) and $x=10^{-1}$ (right), and at the initial scale $Q_0 = 1.64$ GeV. Black curves refer to the most representative replica \#11.}
\label{fig:T_odd_TMDs_f}
\end{figure}

In upper panels of Fig.\tref{fig:T_odd_TMDs_f} we present the transverse-momentum dependence of the $p_T$-weighted $[+,+]$ Sivers function for two representative values of the longitudinal fraction, $x = 10^{-1}$ and $x = 10^{-3}$, and at the initial scale $Q_0 = 1.64$ GeV. Corresponding results for the $[+,+]$ linearity function are given in ower panels. By inspecting our plots, it emerges that both the distributions have a non-Gaussian pattern in $\boldsymbol{p}_T^2$, with a large flattening tail at large $\boldsymbol{p}_T^2$-values and a small but nonzero value when $\boldsymbol{p}_T^2 \to 0$, which suggests that in this limit both TMDs diverge at most as $1/|\boldsymbol{p}_T|$. 
At variance with the $T$-even unpolarized and the Boer--Mulders gluon functions (see Fig.~(4) of Ref.\tcite{Bacchetta:2020vty}), the bulk of our $f$-type $T$-odd functions increases when $x$ grows. This suggests that transverse single-spin asymmetries could be less manifest in the low-$x$ regime.
We remark, however, that our results could change even radically when the full-vertex calculation will become available.

\section{Conclusions and prospects}
\label{summary}

We have enhanced our spectator-model framework by performing preliminary calculation of two $f$-type $T$-odd gluon TMDs: the Sivers and the linarity functions. The full calculation of all the $T$-odd gluon TMDs, including the $d$-type ones is underway. They can serve as a useful guidance to shed light on gluon-TMD dynamics at new-generation particle colliders and experiments, such as the \emph{Electron-Ion Collider}~(EIC)~\cite{AbdulKhalek:2021gbh}, NICA-SPD~\cite{Arbuzov:2020cqg}, the \emph{High-Luminosity Large Hadron Collider} (HL-LHC)~\cite{Chapon:2020heu}, and the \emph{Forward Physics Facility} (FPF)~\cite{Anchordoqui:2021ghd}.

\end{document}